\documentclass{article}
\usepackage{graphicx}
\usepackage{amsmath}

\input{tcilatex}

\begin{document}

\author{Camelia Prodan, Emil Prodan\thanks{%
e-mail: prodan@shasta.phys.uh.edu} \\
University of Houston, 4800 Calhoun Rd, Houston TX 77204-5508}
\title{The dielectric behavior of the living cell suspensions}
\date{08.04.1998}
\maketitle

\begin{abstract}
In the limit of small concentrations and weak applied electric fields, the
dielectric permittivity of suspensions of arbitrarily shaped, shelled and
charged particles is calculated. It is proved that the dielectric behavior
at low frequencies is dominated by the effects of the diffusion of the free
charges on the shell surfaces. Our theoretical formula is valid in the low
range of frequencies ($\alpha $ dispersion) as well as in the high range of
frequencies ($\beta $ dispersion). Will result that one can measure the
membrane electrical potential by a simple investigation of the living cell
suspension dielectric properties.
\end{abstract}

\section{Introduction}

The development of quantitative methods of characterizing the living matter
represents the subject of many researches of our days. The study of the
electric (dielectric) properties of biological systems belongs to this kind
of programs. A simple measurement of the dielectric permittivity of a living
cells suspension provides us with an important set of data which can be used
to describe the living matter \cite{Bo,Sc}. There are many factors which may
influence the dielectric behavior of the biological materials: structure,
molecular dipoles orientation, self interactions, surface conductance,
diffusion, membrane transport processes etc. All these factors influence one
each other and it is hard (if not impossible) to separate the effect of a
single one. However, some effects are dominant in certain ranges of
frequencies and certain conditions. For example, in the $\beta $-dispersion
range (4 MHz-8 MHz), the dielectric behavior of the living cell suspensions
is mainly influenced by the dielectric structure. At low frequencies, it is
dominated by the diffusion effects of the free charges accumulated around
the cell membrane \cite{Gh1}. Also, the orientation of molecular dipoles and
membrane transport processes become relevant when strong electric fields are
applied. For a survey of the field and an extended bibliography one can
consult \cite{Gi}.

The electric activity, in particular, the existence of the membrane
potential is one of the simplest factor which allows us to discern between a
living and a nonliving cell. We will show in this paper its influence on the
dielectric properties of biological materials. Our model do not include
rotational effects or self interactions, so one expects that our results to
be valid only for diluted suspensions and weak electric fields. We like to
think that our work is complementary to the works done in electrophoresis,
where the applied electric fields are very strong and the object of
investigations is the induced membrane potential \cite{Gr}. In this paper,
the applied electric fields are weak and the object of investigation is the
influence of the membrane potential.

We calculate the dielectric permittivity of suspensions of shelled,
arbitrarily shaped particles with a superficial distribution of free charges
on the two faces of the shell, in the limit of small concentrations and weak
applied electric fields. The method used here is similar with that of \cite
{Be}, the basic tools being the functional calculus and the spectral
decomposition of the operators. For the real situation, the distributions of
free charges are rather spatial than superficial. For usual values of the
(3-D) diffusion constants, they have the support in the immediate vicinity
of the membrane (the distributions are practically zero for the distances
larger than 10$^{-9}$ m) and this allows us to treat them as superficial
distributions. Our numerical application will reveal the dependence of the
dielectric permittivity on the membrane electrical potential, mobility of
the free distribution of charges and particle geometry. All these factors
are equally important in the low range of frequency. It turns out that the
consideration of non-spherical geometries (especially non-convex geometries)
may play an important role in some situations. For example, if one wants to
investigate the living cell cycles by dielectric measurements on
synchronized cell suspensions, then definitely, near the subdivision point,
the shape effects are considerable \cite{Gh3}.

We don't know yet how to relate our results with the electrorotation theory 
\cite{Gi2} or how to combine them. We consider that is premature to compare
our theoretical results with the experimental results obtained by
electrorotation, which, at this time, seems to be the only accurate
experimental data in the low range of frequencies. This is mainly because of
different regimes of the applied electric field and concentrations.

\section{The general analysis}

From now, the perturbation of the external electric field (which will be
called the excitation), due to the presence of a particle, will be called
the response of that particle to the excitation. It is known that the
Lorentz method of dielectric permittivity calculus works only if the
response of the suspension constituents is linear in respect to the
excitation. When diffusion effects are considered, this fact is no longer
true. This is the reason we start our analysis with the response of a single
particle to an external excitation.

Let us consider a particle with two dielectric phases (shelled particle),
placed in a spacial homogeneous and temporal oscillating electric field. In
addition, we consider that we have a superficial charge distribution on the
shell faces. Laplace equation, satisfied by the electrical potential, has to
be completed with the boundary conditions on the separation surfaces of the
dielectric phases. These conditions will be deducted from charge
conservation equation. If one considers a surface $\Sigma $ which separates
two dielectric mediums, $\mathcal{D}_{\pm }$, with a free charge $\rho $
distributed on it, then the electrical current density is formed by a volume
one, given by $\overset{\rightarrow }{j^{\pm }}=\sigma ^{\pm }\cdot \overset{%
\rightarrow }{E}$ and a singular density, $\overset{\rightarrow }{j}_{sg}$,
having the support on the interface $\Sigma $. Denoting with $n$ the net
charge concentration (note that a part of charges are due to different
conductivities $\sigma ^{+}$, $\sigma ^{-}$), the continuity equation for a
domain $\mathcal{D}$, centered on the surface, is 
\begin{equation}
-\frac{d}{dt}\underset{\mathcal{D}}{\int }ndv=\underset{\partial \mathcal{D}%
}{\oint }\overset{\rightarrow }{j}_{total}d\overset{\rightarrow }{S}=%
\underset{\partial \mathcal{D}}{\oint }\overset{\rightarrow }{j}_{vol}d%
\overset{\rightarrow }{S}+\underset{\Gamma }{\oint }\overset{\rightarrow }{j}%
_{sg}d\overset{\rightarrow }{\Gamma }\text{,}
\end{equation}
where $\Gamma =\partial \mathcal{D}\cap \Sigma $. Using the Maxwell
equation, $n=div\vec{D}$, it follows 
\begin{equation}
\underset{\partial \mathcal{D}}{\oint }(\sigma \cdot \overset{\rightarrow }{E%
}+\frac{\partial }{\partial t}\overset{\rightarrow }{D})\cdot d\overset{%
\rightarrow }{S}=-\underset{\Gamma }{\oint }\overset{\rightarrow }{j}_{sg}d%
\overset{\rightarrow }{\Gamma }\text{.}
\end{equation}
The singular current is the only cause of the superficial distribution
variation 
\begin{equation}
-\frac{d}{dt}\underset{\mathcal{D}\cap \Sigma }{\int }\rho \cdot dS=%
\underset{\Gamma }{\oint }\overset{\rightarrow }{j}_{sg}d\overset{%
\rightarrow }{\Gamma }
\end{equation}
Finally, the passing equation of the electrical field, in the integral form,
is 
\begin{equation}
\underset{\partial \mathcal{D}}{\oint }(\sigma \cdot \overset{\rightarrow }{E%
}+\frac{\partial }{\partial t}\overset{\rightarrow }{D})\cdot d\overset{%
\rightarrow }{S}=\frac{d}{dt}\underset{\mathcal{D}\cap \Sigma }{\int }\rho
\,dS\text{,}
\end{equation}
and in the differential form is 
\begin{equation}
\vec{n}(\sigma ^{+}\cdot \overset{\rightarrow }{E^{+}}+\frac{\partial }{%
\partial t}\overset{\rightarrow }{D^{+}})-\vec{n}(\sigma ^{-}\cdot \overset{%
\rightarrow }{E^{-}}+\frac{\partial }{\partial t}\overset{\rightarrow }{D^{-}%
})=\frac{\partial \rho }{\partial t}\text{,}
\end{equation}
where $\vec{n}$ represents the normal at the interface. For shelled
particles, the complete system of equations will be 
\begin{equation}
\left\{ 
\begin{array}{l}
\Delta \Phi =0\text{ ; }x\in \Re ^{3}\backslash (\Sigma _{1}\cup \Sigma _{2})
\\ 
\overset{\rightarrow }{n}(\sigma ^{+}\cdot \overset{\rightarrow }{E}^{+}+%
\dfrac{\partial }{\partial t}\overset{\rightarrow }{D}^{+})-\overset{%
\rightarrow }{n}(\sigma ^{-}\cdot \overset{\rightarrow }{E}^{-}+\dfrac{%
\partial }{\partial t}\overset{\rightarrow }{D}^{-})=\dfrac{\partial \rho }{%
\partial t} \\ 
\text{ }x\in \Sigma _{1}\cup \Sigma _{2} \\ 
div_{_{\Sigma }}\overset{\rightarrow }{j}_{sg}=-\dfrac{\partial \rho }{%
\partial t}\text{; }x\in \Sigma _{1}\cup \Sigma _{2} \\ 
\overset{\rightarrow }{E}\longrightarrow \overset{\rightarrow }{E_{0}}\cdot
\exp (j\omega _{0}t)\text{ as }\left| \overset{\rightarrow }{x}\right|
\rightarrow \infty
\end{array}
\right.
\end{equation}
where $\Sigma _{1}$, $\Sigma _{2}$ are the external respective the internal
face of the shell and $\overset{\rightarrow }{E}_{0}\exp \left( j\omega
_{0}t\right) $ is the external field.

The temporal Fourier decomposition leads to the following form of the
equations and boundary conditions: 
\begin{equation}
\left\{ 
\begin{array}{l}
\Delta \Phi =0\text{ ; }x\in \Re ^{3}\backslash (\Sigma _{1}\cup \Sigma _{2})
\\ 
(\sigma ^{+}+j\omega \varepsilon ^{+})\overset{\rightarrow }{n}\cdot \vec{E}%
^{+}-(\sigma ^{-}+j\omega \varepsilon ^{-})\overset{\rightarrow }{n}\cdot 
\vec{E}^{-}=j\omega \rho \text{,} \\ 
\text{ }x\in \Sigma _{1}\cup \Sigma _{2} \\ 
div_{_{\Sigma }}\overset{\rightarrow }{j}_{sg}=-j\omega \rho \text{; }x\in
\Sigma 1\cup \Sigma _{2} \\ 
\overset{\rightarrow }{E}\longrightarrow \left\{ 
\begin{array}{c}
\overset{\rightarrow }{E_{0}}\text{ ; }\omega =\omega _{0} \\ 
0\text{; }\omega \neq \omega _{0}
\end{array}
\right. \text{ as }\left| \overset{\rightarrow }{x}\right| \rightarrow
\infty \text{.}
\end{array}
\right.
\end{equation}
For $\omega \neq 0$, the equation $div_{_{\Sigma }}\overset{\rightarrow }{j}%
_{sg}=-j\omega \rho $ imposes 
\begin{equation}
\underset{\Sigma _{1,2}}{\int }\rho (\omega )dS=-\frac{1}{j\omega }\underset{%
\Sigma _{1,2}}{\int }div_{_{\Sigma }}\overset{\rightarrow }{j}_{sg}dS=0\text{%
.}
\end{equation}
This means that for $\omega \neq \omega _{0}$ and $\omega \neq 0$, we obtain
trivial boundary conditions. So the system has a trivial solution. For $%
\omega =0$, the above equation cannot be written. It is being replaced with
the condition 
\begin{equation}
\underset{\Sigma _{1,2}}{\int }\rho (0)dS=\pm Q\text{,}
\end{equation}
where $\pm Q$ represent the total free charge distributed on the two
interfaces. This leads to nontrivial boundary conditions and consequently to
a nontrivial solution. In consequence, the solution of the system has the
following general form: 
\begin{equation}
\left\{ 
\begin{array}{c}
\Phi (x,t)=\Phi _{0}(x)+\Phi (x)\cdot \exp (j\omega _{0}t) \\ 
\rho (x,t)=\rho _{0}(x)+\rho (x)\cdot \exp (j\omega _{0}t)
\end{array}
\right.
\end{equation}
The components $\Phi _{0}$ and $\rho _{0}$ represent the solution of the
system for zero external excitation and will represent the equilibrium of
the system. The equilibrium will be discussed in the next section. The
components $\Phi $ and $\rho $ are solutions of the following system: 
\begin{equation}
\left\{ 
\begin{array}{l}
\Delta \Phi =0\text{ ; }x\in \Re ^{3}\backslash (\Sigma _{1}\cup \Sigma _{2})
\\ 
(\sigma ^{+}+j\omega \varepsilon ^{+})\dfrac{\partial \Phi ^{+}}{\partial 
\overset{\rightarrow }{n}}-(\sigma ^{-}+j\omega \varepsilon ^{-})\dfrac{%
\partial \Phi ^{-}}{\partial \overset{\rightarrow }{n}}=j\omega \rho \text{; 
}x\in \Sigma _{1}\cup \Sigma _{2} \\ 
div_{_{\Sigma }}\overset{\rightarrow }{j}_{sg}=-j\omega \rho \text{; }x\in
\Sigma _{1}\cup \Sigma _{2} \\ 
\overset{\rightarrow }{E}\longrightarrow \overset{\rightarrow }{E_{0}}\text{
as }\left| \overset{\rightarrow }{x}\right| \rightarrow \infty \text{.}
\end{array}
\right.
\end{equation}
where the pulsation of the excitation was re-denoted with $\omega $.
Introducing the expression of the singular current 
\begin{equation}
\overset{\rightarrow }{j}_{sg}=\overset{\rightarrow }{j}_{conduction}+%
\overset{\rightarrow }{j}_{diffusion}=-\gamma _{i}\cdot \vec{\nabla}_{\Sigma
_{i}}\Phi -D_{i}\vec{\nabla}_{\Sigma _{i}}\rho _{i}\text{, }i=1,2\text{,}
\end{equation}
and working with the complex electric permittivity $\varepsilon ^{\ast
}=\varepsilon +\sigma /\left( j\omega \right) $, we can write 
\begin{equation}
\left\{ 
\begin{array}{l}
\Delta \Phi =0\text{ ; }x\in \Re ^{3}\backslash (\Sigma _{1}\cup \Sigma _{2})
\\ 
\varepsilon _{i-1}^{\ast }\dfrac{\partial \Phi ^{+}}{\partial \overset{%
\rightarrow }{n}}-\varepsilon _{i}^{\ast }\dfrac{\partial \Phi ^{-}}{%
\partial \overset{\rightarrow }{n}}=\rho _{i}\text{; }x\in \Sigma _{i} \\ 
div_{_{\Sigma _{i}}}[-\gamma _{i}\cdot \vec{\nabla}_{_{\Sigma _{i}}}\Phi
-D_{i}\vec{\nabla}_{_{\Sigma _{i}}}\rho _{i}]=-j\omega \rho _{i}\text{; }%
x\in \Sigma _{i} \\ 
\overset{\rightarrow }{E}\longrightarrow \overset{\rightarrow }{E_{0}}\text{
as }\left| \overset{\rightarrow }{x}\right| \rightarrow \infty \text{.}
\end{array}
\right.
\end{equation}
where the index $\Sigma $ means that the operators are calculated on the
surface. The system is not linear, because the conductivities of the
superficial charges, $\gamma _{i}$, depend on $\rho $. We will consider in
the following that we are in the limit of weak external electric fields so
we can consider that the conductivities of the free charges are given by the
equilibrium configuration. Thus the system becomes linear. Now, the total
electric field will be the sum of the electrical fields of the equilibrium
configuration and of the perturbation, $\overset{\rightarrow }{E}_{total}=%
\overset{\rightarrow }{E}_{e}+\overset{\rightarrow }{E}$. In the weak fields
approximation, the average on the different orientations of the particle
leads to the following value of the electric field inside of the particle: 
\begin{equation}
\left\langle \overset{\rightarrow }{E}\right\rangle _{orientations}=\overset{%
\rightarrow }{E}_{0}\cdot \frac{1}{4\pi }\underset{\Omega _{\overset{%
\rightarrow }{N}}}{\int }\overset{\rightarrow }{E}_{\overset{\rightarrow }{N}%
}\cdot \overset{\rightarrow }{N}\ d\Omega _{\overset{\rightarrow }{N}}\text{,%
}
\end{equation}
where $\overset{\rightarrow }{E}_{\overset{\rightarrow }{N}}$ is the
electric field (only the perturbation part) inside of the particle, when the
external electric field, $\overset{\rightarrow }{N}$, is of norm one.
Further, this average can be calculate with the formula 
\begin{equation}
\left\langle \overset{\rightarrow }{E}\right\rangle _{orientations}=\frac{1}{%
3}\left( \underset{i}{\sum }\overset{\rightarrow }{N}_{i}\cdot \overset{%
\rightarrow }{E}_{\overset{\rightarrow }{N}_{i}}\right) \overset{\rightarrow 
}{E}_{0}\text{,}
\end{equation}
where $i$ designates three orthogonal directions. The average electric field
is proportional with the excitation. This allows us to use the Lorenz
formula 
\begin{equation}
\varepsilon _{sus}=\varepsilon _{o}\left( 1+\frac{p\alpha }{1-\frac{p\alpha 
}{3}}\right)
\end{equation}
for the suspension dielectric permittivity calculus. Here $p$ is the volume
concentration of the suspension and $\alpha $ is the polarization of the
particles 
\begin{equation}
\alpha =\frac{1}{VE_{0}}\underset{V}{\int }dv\cdot \frac{\varepsilon
-\varepsilon _{o}}{\varepsilon _{o}}\left\langle \overset{\rightarrow }{E}%
\right\rangle _{orientations}
\end{equation}
where $V$ is the particle volume, $\varepsilon _{o}$ is the dielectric
permittivity of the exterior medium and $\varepsilon $ is the dielectric
permittivity of the particle.

\section{The equilibrium}

In the following the primitivities and the conductivities will be denoted
by: $\varepsilon _{0}$, $\sigma _{0}$ for the exterior, $\varepsilon _{1}$, $%
\sigma _{1}$ for the membrane, $\varepsilon _{2}$, $\sigma _{2}$ for the
interior. The system which establish the equilibrium configuration is
obtained from the general system by cancelling the temporal derivatives: 
\begin{equation}
\left\{ 
\begin{array}{l}
\Delta \Phi _{e}=0\text{ ; }x\in \Re ^{3}\backslash (\Sigma _{1}\cup \Sigma
_{2}) \\ 
\sigma _{i-1}\dfrac{\partial \Phi _{e}^{+}}{\partial \overset{\rightarrow }{n%
}}\mid _{\Sigma _{i}}=\sigma _{i}\dfrac{\partial \Phi _{e}^{-}}{\partial 
\overset{\rightarrow }{n}}\mid _{\Sigma _{i}}\text{; }i=1,2 \\ 
div_{_{\Sigma _{i}}}\overset{\rightarrow }{j}_{sg}=0 \\ 
\int_{\Sigma _{1,2}}\rho _{0}dS=\pm Q\text{.}
\end{array}
\right.
\end{equation}
If we consider the conductivity of the membrane, $\sigma _{1}$, equal with
zero, we obtain trivial Neumann condition for the electric potential, inside
and outside of the particle. This imposes constant electrical potential in
the two regions. The distribution of charges on the two interfaces, $\tau
_{1,2}$, can be calculated by applying the Gauss law: 
\begin{equation}
\tau _{1}=\varepsilon _{1}\dfrac{\partial \Phi ^{-}}{\partial \overset{%
\rightarrow }{n}}\mid _{\Sigma _{1}};\ \,\tau _{2}=\varepsilon _{1}\dfrac{%
\partial \Phi ^{+}}{\partial \overset{\rightarrow }{n}}\mid _{\Sigma _{2}}
\end{equation}
where $\tau $ includes both, the free charge $\rho $ and the charges which
are accumulated on the interfaces due to the conduction currents. The
equations for free charges are 
\begin{equation}
div[\gamma _{i}\cdot \overset{\rightarrow }{E}_{t}-D\nabla \rho
_{0_{i}}]=0\Longrightarrow D\Delta \rho _{0_{i}}=0\Longleftrightarrow \rho
_{0_{i}}=const.\text{,}
\end{equation}
where $\overset{\rightarrow }{E}_{t}$ is the tangent electric field to $%
\Sigma _{1,2}$. The link between $\rho _{0}$ and $Q$ results by the reason
that the inside region is not an electric charges reservoir, which means
that the total charge on the interfaces is given only by the free charges.
Thus 
\begin{equation}
\pm Q=\underset{\Sigma _{1,2}}{\int }\rho _{0_{1,2}}dS=\rho _{0_{1,2}}\cdot
S_{1,2}\ ;\ Q=C\cdot \Delta V_{0}
\end{equation}
where $\Delta V_{0}$ is the electric potential difference between the two
surfaces (membrane potential) and $C$ is the capacity of a condenser having
the geometry of the shell and dielectric permittivity equal with $%
\varepsilon _{1}$. The values of the $\rho _{0_{i}}$ fix the electric
conductivity on the two interfaces 
\begin{equation}
\gamma _{1,2}=\rho _{0_{1,2}}\cdot u_{1,2}=\pm u_{1,2}\frac{C}{S_{1,2}}\cdot
\Delta V_{0}\text{,}
\end{equation}
where $u_{i}$, $i=1$, $2$ are the charge mobilities on the two interfaces.

It was seen that, in the calculus of the dielectric permittivity, only the
perturbation part of the electric field is important. So the only influence
of the membrane electric potential on the dispersions curves of the
dielectric permittivity comes from the above formula.

\section{The effective calculus of the polarization}

The charges on the second interface lie behind the shell, so the external
electric field will have a smaller influence on them than on the charges of
the first interface. In the same time, it is an experimental fact that the
mobility of the inside charges is much smaller than that of external
charges. Both reasons justify the idea of considering the free charges of
the inside interface to be fixed (i.e. zero mobility). With this
simplification, the value of $\rho _{2}$ is fixed at the equilibrium value.
Now we have inhomogeneous particles with free electric charges only on their
surface.

\subsection{The equivalence between an inhomogeneous particle and a
homogeneous one}

Let us consider first the situation of a homogeneous particle, with
dielectric permittivity $\varepsilon $ and conductivity $\sigma $, placed in
the electric field $\overset{\rightarrow }{E}_{0}e^{j\omega t}$. If we use
the expression of the simple layer for the electric potential 
\begin{equation}
\Phi (\overset{\rightarrow }{x})=-\overset{\rightarrow }{x}\cdot \overset{%
\rightarrow }{E_{0}}+\frac{1}{4\pi }\underset{y\in \Sigma }{\cdot \int }%
\frac{\mu (\overset{\rightarrow }{y})}{\left| \overset{\rightarrow }{x}-%
\overset{\rightarrow }{y}\right| }dS_{y}\text{,}
\end{equation}
the passing equation for the electric field through the surface of the
particle, $\Sigma $, becomes the following integral equation for the charge
distribution $\mu $: 
\begin{equation}
\frac{1}{2\lambda }\mu (\overset{\rightarrow }{x})-\hat{E}[\mu ](\overset{%
\rightarrow }{x})=\overset{\rightarrow }{n}\cdot \overset{\rightarrow }{E_{0}%
}\text{,}
\end{equation}
where $\lambda =\dfrac{\varepsilon ^{\ast }-\varepsilon _{o}^{\ast }}{%
\varepsilon ^{\ast }+\varepsilon _{o}^{\ast }}$ and $\hat{E}$ is the
operator 
\begin{equation}
\hat{E}\left[ \mu \right] \left( \overset{\rightarrow }{x}\right) =\frac{1}{%
4\pi }\underset{y\in \Sigma }{\cdot \int }\frac{(\overset{\rightarrow }{x}-%
\overset{\rightarrow }{y})\cdot \overset{\rightarrow }{n_{x}}}{\left| 
\overset{\rightarrow }{x}-\overset{\rightarrow }{y}\right| ^{3}}\mu (%
\overset{\rightarrow }{y})dS_{y}\text{.}
\end{equation}
Using the spectral decomposition of the $\hat{E}$ operator, the solution of
this equation is 
\begin{equation}
\mu =\underset{n}{\sum }\frac{\lambda }{\frac{1}{2}-\lambda \cdot \chi _{n}}%
\cdot \hat{P}_{n}[\overset{\rightarrow }{n}\cdot \overset{\rightarrow }{E_{0}%
}]\text{,}
\end{equation}
where $\hat{P}_{n}$ is the spectral projector corresponding to the
eigenvalue $\chi _{n}$. The $\hat{E}$ operator is not a symmetric one, so
some precautions are needed. About this operator and the validity of the
above decomposition, one can consult \cite{RS}.

Let us consider now the shelled particle placed in the same electric field.
We consider here that the external surface of the shell is obtained by
expanding the internal surfaces by a factor $\delta >1$. The equivalence
problem between this inhomogeneous dielectric particle and a homogeneous one
can be formulated in the following terms: there exists a distribution of
charges only on the external surface of the shell, $\mu _{e}$, which
provides the true electric field outside of particle. In the dipole
approximation, this distribution is given by the following expression: 
\begin{equation}
\mu _{e}=\underset{n}{\sum }\frac{\lambda _{n}}{\frac{1}{2}-\lambda
_{n}\cdot \chi _{n}}\cdot \hat{P}_{n}[\overset{\rightarrow }{n}\cdot 
\overset{\rightarrow }{E_{0}}]\text{.}
\end{equation}
where $\lambda _{n}$ are given by: 
\begin{equation}
\lambda _{n}=\frac{\varepsilon _{n}^{\ast }-\varepsilon _{o}^{\ast }}{%
\varepsilon _{n}^{\ast }+\varepsilon _{o}^{\ast }}\text{; }\varepsilon
_{n}^{\ast }=\varepsilon _{1}^{\ast }\frac{\chi _{n}+\frac{1}{2}+\delta ^{3}(%
\frac{1}{2\lambda _{2}}-\chi _{n})}{\chi _{n}-\frac{1}{2}+\delta ^{3}(\frac{1%
}{2\lambda _{2}}-\chi _{n})}\text{; }\lambda _{2}=\frac{\varepsilon
_{2}^{\ast }-\varepsilon _{1}^{\ast }}{\varepsilon _{2}^{\ast }+\varepsilon
_{1}^{\ast }}\text{.}
\end{equation}
Comparing with the formula for the homogeneous particle, we conclude that
the shelled particle is equivalent to a homogeneous particle which has the
same geometry and volume. The equivalation relation is given by 
\begin{equation}
\varepsilon ^{\ast }\rightarrow \left\{ \varepsilon _{n}^{\ast }\right\} _{n}%
\text{, or }\varepsilon ^{\ast }=\sum_{n}\varepsilon ^{\ast }\hat{P}%
_{n}\rightarrow \sum_{n}\varepsilon _{n}^{\ast }\hat{P}_{n}\text{.}
\end{equation}
In the particular case of the sphere, the above formula is identical to the
existent formula \cite{Pa}. We will give in the following a sketch of the
proof. For the shelled particle, the single layer expression of the
potential is 
\begin{equation*}
\Phi \left( x\right) =-\vec{x}E_{0}+
\end{equation*}
\begin{equation}
\frac{1}{4\pi }\underset{y\in \Sigma _{1}}{\int }\frac{\mu _{1}(\overset{%
\rightarrow }{y})}{\left| \overset{\rightarrow }{x}-\overset{\rightarrow }{y}%
\right| }dS_{y}+\frac{1}{4\pi }\underset{y\in \Sigma _{2}}{\int }\frac{\mu
_{2}(\overset{\rightarrow }{y})}{\left| \overset{\rightarrow }{x}-\overset{%
\rightarrow }{y}\right| }dS_{y}\text{.}
\end{equation}
The electric potential created by $\mu _{e}$ will be denoted with $\Phi _{e}$%
. The potential $\Phi _{e}$ is equal outside of particle with the exact
potential, $\Phi $, only if the two potentials fulfill the same Neuman
boundary conditions: $\dfrac{\partial \Phi _{e}^{+}}{\partial \vec{n}}=%
\dfrac{\partial \Phi ^{+}}{\partial \vec{n}}$. Using the single layer
expression, the above condition can be written as 
\begin{equation}
\frac{1}{2}\mu _{e}-\hat{E}\left[ \mu _{e}\right] =\frac{1}{2}\mu _{1}-\hat{E%
}_{11}\left[ \mu _{1}\right] -\hat{E}_{12}\left[ \mu _{2}\right] \text{,}
\end{equation}
where 
\begin{equation}
\hat{E}_{ij}\left[ \mu _{j}\right] =\frac{1}{4\pi }\underset{y\in \Sigma _{j}%
}{\cdot \int }\frac{(\overset{\rightarrow }{x}-\overset{\rightarrow }{y}%
)\cdot \overset{\rightarrow }{n_{x}}}{\left| \overset{\rightarrow }{x}-%
\overset{\rightarrow }{y}\right| ^{3}}\cdot \mu _{j}(\overset{\rightarrow }{y%
})dS_{y}\text{; }x\in \Sigma _{i}\text{.}
\end{equation}
The passing equations of the exact electric field, $\Phi $, through the
shell surfaces take the form 
\begin{equation}
\left\{ 
\begin{tabular}{l}
$\frac{1}{2\lambda _{1}}\mu _{1}-\hat{E}_{11}\left[ \mu _{1}\right] -\hat{E}%
_{12}\left[ \mu _{2}\right] =\vec{n}\vec{E}_{0}$ \\ 
$\frac{1}{2\lambda _{2}}\mu _{2}-\hat{E}_{21}\left[ \mu _{1}\right] -\hat{E}%
_{22}\left[ \mu _{2}\right] =\vec{n}\vec{E}_{0}$%
\end{tabular}
\right.
\end{equation}
where $\lambda _{i}=\dfrac{\varepsilon _{i}^{\ast }-\varepsilon _{i-1}^{\ast
}}{\varepsilon _{i}^{\ast }+\varepsilon _{i-1}^{\ast }}$. Here comes the
dipole approximation: 
\begin{equation}
\left\{ 
\begin{tabular}{l}
\begin{tabular}{l}
$\hat{E}_{12}\left[ \mu _{2}\right] \approx \frac{1}{\delta ^{3}}\hat{E}_{22}%
\left[ \mu _{2}\right] $ \\ 
$\hat{E}_{21}\left[ \mu _{1}\right] \approx \hat{E}_{11}\left[ \mu _{1}%
\right] $%
\end{tabular}
\end{tabular}
\right.
\end{equation}
which says that the normal component of the electric field created by the
charge distribution $\mu _{2}$ on the surfaces $\Sigma _{1}$ has the same
angular dependence as on the surfaces $\Sigma _{2}$ but the strength of the
field is diminished by a factor $\delta ^{-3}$, and that the normal
component of the electric field created by the distribution $\mu _{1}$ on
the surface $\Sigma _{2}$ is equal with the normal component on the surface $%
\Sigma _{1}$. With the observation that the operator $\hat{E}$ is scale
invariant, the above equations become 
\begin{equation}
\left\{ 
\begin{tabular}{l}
$\frac{1}{2\lambda _{1}}\mu _{1}-\hat{E}\left[ \mu _{1}\right] -\frac{1}{%
\delta ^{3}}\hat{E}\left[ \mu _{2}\right] =\vec{n}\vec{E}_{0}$ \\ 
$\frac{1}{2\lambda _{2}}\mu _{2}-\hat{E}\left[ \mu _{1}\right] -\hat{E}\left[
\mu _{2}\right] =\vec{n}\vec{E}_{0}$%
\end{tabular}
\right.
\end{equation}
Using again the spectral decomposition of the operator $\hat{E}$ and the
link (31) between $\mu _{e}$ and $\mu _{1}$, $\mu _{2}$, the desired form of
the distribution $\mu _{e}$ follows immediately. One can see that the method
can be extended to the multi-shelled particles. In this way, the results of 
\cite{As1} can be extended to arbitrary geometries.

Now, we want to consider the situation when the free distribution of charges 
$\rho $ is present. For a homogeneous dielectric particle with dielectric
permittivity $\varepsilon $ and conductivity $\sigma $ which has a free
charge on its surface, the equations are 
\begin{equation}
\left\{ 
\begin{array}{l}
\varepsilon _{o}^{\ast }\left( \overset{\rightarrow }{n}\cdot \overset{%
\rightarrow }{E_{0}}+\frac{1}{2}\mu +E[\mu ]\right) -\varepsilon ^{\ast
}\left( \overset{\rightarrow }{n}\cdot \overset{\rightarrow }{E_{0}}-\frac{1%
}{2}\mu +E[\mu ]\right) =\rho \\ 
\left[ \Delta _{\Sigma }+\frac{j\omega }{D}\right] \rho =-\frac{\gamma }{D}%
\Delta _{\Sigma }\Phi
\end{array}
\right.
\end{equation}
where $\Delta _{\Sigma }$ is the Laplace operator on the particle surface.
Here we have continued the idea of the second section by introducing the
simple layer expression of the electric potential. We can pass to the
situation of a shelled particle by using the equivalence relation $%
\varepsilon ^{\ast }\rightarrow \sum_{n}\varepsilon _{n}^{\ast }\hat{P}_{n}$%
. (Here we use our assumption that the distribution of the free charges of
the internal shell surface can be considered immobile). The equations become 
\begin{equation}
\left\{ 
\begin{array}{l}
\varepsilon _{o}^{\ast }\left( \overset{\rightarrow }{n}\cdot \overset{%
\rightarrow }{E_{0}}+\frac{1}{2}\mu +E[\mu ]\right) \\ 
-\left( \underset{n}{\sum }\varepsilon _{n}^{\ast }\cdot \hat{P}_{n}\right)
\left( \overset{\rightarrow }{n}\cdot \overset{\rightarrow }{E_{0}}-\frac{1}{%
2}\mu +E[\mu ]\right) =\rho \\ 
\left( \Delta _{\Sigma _{1}}+\frac{j\omega }{D}\right) \rho =-\frac{\gamma }{%
D}\Delta _{\Sigma _{1}}\Phi
\end{array}
\right.
\end{equation}
Here, $\rho $ denotes the free distribution of charges of the external shell
surface. One observes that the single layer distribution can be decomposed
into $\mu =\mu _{\beta }+\mu _{\alpha }$, where $\mu _{\beta }$ solves the
equations for $\rho =0$. It turns out that $\mu _{\beta }$ is practically
constant at low frequencies and that $\mu _{\alpha }$ is practically zero at
high frequencies. In consequence, the $\mu _{\beta }$ distribution is
responsible for the behavior at high frequencies (the range of the beta
dispersion) and the addition of $\mu _{\alpha }$ will give us the behavior
at low frequencies (the range of alpha dispersion). The equations for $\mu
_{\alpha }$ and $\rho $ will be: 
\begin{equation}
\left\{ 
\begin{array}{l}
\underset{n}{\sum }\left( \frac{1}{2}(\varepsilon _{n}^{\ast }+\varepsilon
_{o}^{\ast })-\chi _{n}(\varepsilon _{n}^{\ast }-\varepsilon _{o}^{\ast
})\right) \hat{P}_{n}\,\mu _{\alpha }=\rho \\ 
\left( \Delta +\frac{j\omega }{D}\right) \rho =-\frac{\gamma }{D}\Delta \Phi
_{\alpha }-\frac{\gamma }{D}\Delta \Phi _{\beta }
\end{array}
\right.
\end{equation}
where $\Phi _{\beta }$ ( $\Phi _{\alpha }$) represents the electric
potential generate by the $\mu _{\beta }$ ($\mu _{\alpha }$) distribution.
The spectral decomposition of the Laplace operator leads to the following
equation for $\mu _{\alpha }$: 
\begin{equation}
\begin{array}{l}
\underset{n}{\sum }\left( \frac{1}{2}(\varepsilon _{n}^{\ast }+\varepsilon
_{o}^{\ast })-\chi _{n}(\varepsilon _{n}^{\ast }-\varepsilon _{o}^{\ast
})\right) \hat{P}_{n}\,\mu _{\alpha } \\ 
+\frac{\gamma }{D}\underset{i}{\sum }\dfrac{\xi _{i}}{\xi _{i}+j\omega }\hat{%
G}_{i}\tilde{\Phi}=-\frac{\gamma }{D}\underset{i}{\sum }\dfrac{\xi _{i}}{\xi
_{i}+j\omega }\hat{G}_{i}\Phi \text{,}
\end{array}
\end{equation}
where $\hat{G}_{i}$ is the spectral projector of the Laplace operator
corresponding to the eigenvalue $\xi _{i}$. Denoting with $\hat{K}$ the
operator 
\begin{equation}
(\hat{K}\mu )(\vec{x})=\frac{1}{4\pi }\underset{y\in \Sigma }{\int }\frac{%
\mu (\vec{y})}{\left| \vec{x}-\vec{y}\right| }dS_{y}\text{,}
\end{equation}
the equation takes the form: 
\begin{equation}
\begin{array}{l}
\left\{ \frac{D}{\gamma }\underset{n}{\sum }\left( \frac{1}{2}(\varepsilon
_{n}^{\ast }+\varepsilon _{o}^{\ast })-\chi _{n}(\varepsilon _{n}^{\ast
}-\varepsilon _{o}^{\ast })\right) \hat{P}_{n}\right. \\ 
\left. +\underset{i}{\sum }\dfrac{\xi _{i}}{\xi _{i}+j\omega }\hat{G}%
_{i}\circ \hat{K}\right\} \mu _{\alpha }=-\underset{i}{\sum }\dfrac{\xi _{i}%
}{\xi _{i}+j\omega }\hat{G}_{i}\circ \left( -\vec{x}\cdot \vec{N}+\hat{K}\mu
_{\beta }\right) \text{.}
\end{array}
\end{equation}
We will discuss later our method of solving this equation.

\subsection{The formula for polarization}

We can use the idea of equivalence to find a compact formula for
polarization. For this, we have to transform the volume integral which
appears in the definition of $\alpha $, 
\begin{equation}
\alpha =\frac{1}{V}\int d\Omega _{\overset{\rightarrow }{N}}\underset{V}{%
\int }dv\cdot \frac{\varepsilon -\varepsilon _{o}}{\varepsilon _{o}}\overset{%
\rightarrow }{N}\cdot \overset{\rightarrow }{E}\text{,}
\end{equation}
into an integral on the external side of the particle surface where we can
use the equivalence. In the above formula, $\varepsilon $ depends on the
point, and $\overset{\rightarrow }{E}$ is the exact field inside of the
particle, prodeced by the excitation $\vec{N}$,\ of norm one. We will
calculate the term as follow: 
\begin{equation}
\begin{array}{l}
\underset{V}{\int }dv\cdot \dfrac{\varepsilon -\varepsilon _{o}}{\varepsilon
_{o}}\overset{\rightarrow }{N}\cdot \overset{\rightarrow }{E}=\underset{V}{%
\int }dv\,\left\{ \dfrac{1}{\varepsilon _{o}}grad\left( \overset{\rightarrow 
}{x}\cdot \overset{\rightarrow }{N}\right) \cdot \overset{\rightarrow }{D}%
+\,div\left( \overset{\rightarrow }{N}\Phi \right) \right\} \\ 
=\dfrac{1}{\varepsilon _{o}}\underset{V}{\int }dv\,div\left( (\overset{%
\rightarrow }{x}\cdot \overset{\rightarrow }{N})\cdot \overset{\rightarrow }{%
D}\right) +\underset{V}{\int }dv\,div\left( \overset{\rightarrow }{N}\Phi
\right) \\ 
=\dfrac{1}{\varepsilon _{o}}\underset{\Sigma ^{-}}{\int }(\overset{%
\rightarrow }{x}\cdot \overset{\rightarrow }{N})\overset{\rightarrow }{D}%
\cdot d\overset{\rightarrow }{S}+\underset{\Sigma ^{-}}{\int }\Phi \overset{%
\rightarrow }{N}\cdot d\overset{\rightarrow }{S}
\end{array}
\end{equation}
where, if it is used the passing relations for $\vec{D}$, $%
D_{n}^{+}-D_{n}^{-}=\rho $ and the continuity of the potential, the
integrals can be processed on the external side of the surface. 
\begin{equation}
\begin{array}{l}
\underset{V}{\int }dv\cdot \dfrac{\varepsilon -\varepsilon _{o}}{\varepsilon
_{o}}\overset{\rightarrow }{N}\cdot \overset{\rightarrow }{E}=\dfrac{1}{%
\varepsilon _{o}}\underset{\Sigma ^{+}}{\int }(\overset{\rightarrow }{x}%
\cdot \overset{\rightarrow }{N})\left( -\rho +D_{n}^{+}\right) \cdot dS \\ 
+\underset{\Sigma ^{+}}{\int }\Phi \overset{\rightarrow }{N}\cdot d\vec{S}=%
\dfrac{1}{\varepsilon _{o}}\underset{\Sigma ^{+}}{\int }(\overset{%
\rightarrow }{x}\cdot \overset{\rightarrow }{N})\left( -\rho +\varepsilon
_{o}E_{n}^{+}\right) \cdot dS+\underset{\Sigma ^{+}}{\int }\Phi \overset{%
\rightarrow }{N}\cdot d\vec{S}\text{.}
\end{array}
\end{equation}
Outside of particle, the electric field is equal to that of the equivalent
homogeneous particle. The normal component of the electric field can be
expressed with the help of the $\hat{E}$ operator: 
\begin{equation}
E_{n}^{+}=-\frac{\partial \Phi ^{+}}{\partial \overset{\rightarrow }{n}}=%
\overset{\rightarrow }{n}\cdot \overset{\rightarrow }{N}+\frac{1}{2}\mu +%
\hat{E}\mu \text{,}
\end{equation}
where $\mu $ is the equivalating charge distribution (solution of (37)),
resulting: 
\begin{equation}
\begin{array}{l}
\underset{V}{\int }dv\cdot \dfrac{\varepsilon -\varepsilon _{o}}{\varepsilon
_{o}}\overset{\rightarrow }{N}\cdot \overset{\rightarrow }{E}=\dfrac{1}{%
\varepsilon _{o}}\underset{\Sigma ^{+}}{\int }(\overset{\rightarrow }{x}%
\cdot \overset{\rightarrow }{N})\left( -\rho +\varepsilon _{o}\left( 
\overset{\rightarrow }{n}\cdot \overset{\rightarrow }{N}+\frac{1}{2}\mu +%
\hat{E}\mu \right) \right) dS \\ 
-\underset{V}{\int }\overset{\rightarrow }{N}\cdot \vec{E}_{e}\ dv=\dfrac{1}{%
\varepsilon _{o}}\underset{\Sigma ^{-}}{\int }(\overset{\rightarrow }{x}%
\cdot \overset{\rightarrow }{N})\left( \underset{n}{\sum }\varepsilon
_{n}^{\ast }\cdot \hat{P}_{n}\right) \left( \overset{\rightarrow }{n}\cdot 
\overset{\rightarrow }{N}-(\frac{1}{2}-\chi _{n})\mu \right) dS \\ 
-\underset{\Sigma ^{-}}{\int }(\overset{\rightarrow }{x}\cdot \overset{%
\rightarrow }{N})\overset{\rightarrow }{E_{e}}d\overset{\rightarrow }{S}
\end{array}
\end{equation}
Having that $\vec{n}\vec{E}_{e}^{-}=\dfrac{\partial \Phi _{e}}{\partial \vec{%
n}}=\overset{\rightarrow }{n}\cdot \overset{\rightarrow }{N}-\frac{1}{2}\mu +%
\hat{E}\left[ \mu \right] $, we can continue 
\begin{equation}
\begin{array}{l}
\underset{V}{\int }dv\cdot \dfrac{\varepsilon -\varepsilon _{o}}{\varepsilon
_{o}}\overset{\rightarrow }{N}\cdot \overset{\rightarrow }{E} \\ 
=\underset{\Sigma }{\int }(\overset{\rightarrow }{x}\cdot \overset{%
\rightarrow }{N})\left( \underset{n}{\sum }\dfrac{(\varepsilon _{n}^{\ast
}-\varepsilon _{o})}{\varepsilon _{o}}\cdot \hat{P}_{n}\right) \left( 
\overset{\rightarrow }{n}\cdot \overset{\rightarrow }{N}-(\frac{1}{2}-\chi
_{n})\mu \right) dS \\ 
=\underset{n}{\sum }\dfrac{(\varepsilon _{n}^{\ast }-\varepsilon _{o})}{%
\varepsilon _{o}}\cdot \underset{\Sigma }{\int }(\overset{\rightarrow }{x}%
\cdot \overset{\rightarrow }{N})\hat{P}_{n}\left( \overset{\rightarrow }{n}%
\cdot \overset{\rightarrow }{N}-(\frac{1}{2}-\chi _{n})\mu \right) dS\text{.}
\end{array}
\end{equation}
If one denotes the scalar product on the $L^{2}(\Sigma )$ functions space by 
$\left\langle \,,\right\rangle $, 
\begin{equation}
\left\langle \varphi ,\phi \right\rangle =\underset{\Sigma }{\int }\bar{%
\varphi}(\overset{\rightarrow }{x})\phi (\overset{\rightarrow }{x})\,dS_{x}%
\text{, }\forall \text{ }\varphi \text{, }\phi \in L^{2}(\Sigma )\text{,}
\end{equation}
we obtain 
\begin{equation}
\begin{array}{l}
\underset{V}{\int }dv\cdot \dfrac{\varepsilon -\varepsilon _{o}}{\varepsilon
_{o}}\overset{\rightarrow }{N}\cdot \overset{\rightarrow }{E}=\underset{n}{%
\sum }\dfrac{(\varepsilon _{n}^{\ast }-\varepsilon _{o}^{\ast })}{%
\varepsilon _{o}^{\ast }}\times \\ 
\left\{ \left\langle \overset{\rightarrow }{x}\cdot \overset{\rightarrow }{N}%
\mid \hat{P}_{n}\mid \overset{\rightarrow }{n}\cdot \overset{\rightarrow }{N}%
\right\rangle -(\frac{1}{2}-\chi _{n})\left\langle \overset{\rightarrow }{x}%
\cdot \overset{\rightarrow }{N}\mid \hat{P}_{n}\mid \mu \right\rangle
\right\}
\end{array}
\end{equation}
Having in view the discussion about the average on the orientations, the
angular integral can be eliminated and the final formula for polarization
becomes 
\begin{equation}
\begin{array}{l}
\alpha =\frac{1}{3V}\underset{n,i}{\sum }\frac{(\varepsilon _{n}^{\ast
}-\varepsilon _{o}^{\ast })}{\varepsilon _{o}^{\ast }}\left\{ \left\langle 
\overset{\rightarrow }{x}\cdot \overset{\rightarrow }{N_{i}}\mid \hat{P}%
_{n}\mid \overset{\rightarrow }{n}\cdot \overset{\rightarrow }{N_{i}}%
\right\rangle \right. \\ 
\left. -(\frac{1}{2}-\chi _{n})\left\langle \overset{\rightarrow }{x}\cdot 
\overset{\rightarrow }{N_{i}}\mid \hat{P}_{n}\mid \mu ^{i}\right\rangle
\right\} \text{,}
\end{array}
\end{equation}
where $i=1,2,3$ denotes three orthogonal directions and $\mu ^{i}$ denotes
the total distribution of the single layer corresponding at the excitation $%
\vec{N}_{i}$. If the free charge is not present, the above formula becomes 
\begin{equation}
\begin{array}{l}
\alpha =\frac{1}{3V}\underset{n,i}{\sum }\dfrac{(\varepsilon _{n}^{\ast
}-\varepsilon _{o}^{\ast })}{\varepsilon _{o}^{\ast }}\left\{ \left\langle 
\overset{\rightarrow }{x}\cdot \overset{\rightarrow }{N_{i}}\mid \hat{P}%
_{n}\mid \overset{\rightarrow }{n}\cdot \overset{\rightarrow }{N_{i}}%
\right\rangle \right. \\ 
\left. -(\frac{1}{2}-\chi _{n})\left\langle \overset{\rightarrow }{x}\cdot 
\overset{\rightarrow }{N_{i}}\mid \frac{\lambda _{n}}{\frac{1}{2}-\chi
_{n}\lambda _{n}}\hat{P}_{n}\mid \overset{\rightarrow }{n}\cdot \overset{%
\rightarrow }{N_{i}}\right\rangle \right\} \\ 
=\frac{1}{3V}\underset{n,i}{\sum }\dfrac{\lambda _{n}}{\frac{1}{2}-\chi
_{n}\lambda _{n}}\left\langle \overset{\rightarrow }{x}\cdot \overset{%
\rightarrow }{N_{i}}\mid \hat{P}_{n}\mid \overset{\rightarrow }{n}\cdot 
\overset{\rightarrow }{N_{i}}\right\rangle \text{.}
\end{array}
\end{equation}
This formula is exact in the high frequencies range and it describes
correctly the $\beta $ dispersion region. The exact formula of $\alpha $
must be completed with the term which comes from $\mu _{\alpha }$%
\begin{equation}
\begin{array}{l}
\alpha =\frac{1}{3V}\underset{n,i}{\sum }\dfrac{\lambda _{n}}{\frac{1}{2}%
-\chi _{n}\lambda _{n}}\left\langle \overset{\rightarrow }{x}\cdot \overset{%
\rightarrow }{N_{i}}\mid \hat{P}_{n}\mid \overset{\rightarrow }{n}\cdot 
\overset{\rightarrow }{N_{i}}\right\rangle \\ 
-\frac{1}{3V}\underset{n,i}{\sum }\dfrac{(\frac{1}{2}-\chi _{n})(\varepsilon
_{n}^{\ast }-\varepsilon _{o}^{\ast })}{\varepsilon _{o}^{\ast }}%
\left\langle \overset{\rightarrow }{x}\cdot \overset{\rightarrow }{N_{i}}%
\mid \hat{P}_{n}\mid \mu _{\alpha }^{i}\right\rangle
\end{array}
\end{equation}
As we already discussed, the last term is practically zero at high
frequencies and becomes dominant at low frequencies.

\section{Numerical results}

First, we will present our method of solving the complicated equation (41)
of $\mu _{\alpha }$. Here, the principal problem is the calculation of the
spectral projectors of $\hat{E}$ and $\Delta _{\Sigma }$ operators. Let us
suppose that our surface is given in the spherical coordinates by $r=r\left(
\theta ,\varphi \right) $. To solve the problem for the $\hat{E}$ operator,
we will choose the following orthonormal basis in $L^{2}\left( \Sigma
\right) $: 
\begin{equation}
\left\{ \mathcal{Y}_{lm}\right\} _{\substack{ m\in \mathbf{Z}  \\ l\geq
\left| m\right| }}\text{; }\mathcal{Y}_{lm}\left( \theta ,\varphi \right) =%
\frac{Y_{lm}\left( \theta ,\varphi \right) }{\sqrt{\sigma \left( \theta
,\varphi \right) }}\text{,}
\end{equation}
where $\left\{ Y_{lm}\right\} _{l,m}$ represents the spherical harmonics and 
$dS=\sigma \left( \theta ,\varphi \right) \sin \left( \theta \right) d\theta
d\varphi $. The first step is the calculation of the matrix elements of the
operator $\hat{E}$. Then, taking acceptable truncated matrices, we will
calculate the approximative spectrum. The approximative projectors will be
given by $\hat{P}_{n}=P_{n}\left( \hat{E}\right) $, where $P_{n}$ are the
polynomials which satisfy $P_{n}\left( \chi _{m}\right) =\delta _{nm}$ for
each eigenvalue $\chi _{m}$. For the operator $\Delta _{\Sigma }$, we have
followed the way presented in \cite{PR2}.

We have chosen the surfaces of figure 1 for our numerical application. As we
already emphasized, $\gamma $ is proportional with the electrical potential
of the membrane and the mobility of the free charges. Because the diffusion
constant is also proportional with the superficial charge mobility, the
quantity $\gamma /D$ is proportional only to the membrane electrical
potential. We will choose different values for the diffusion coefficient, $D$%
, and for the ratio $\gamma /D$, which will give us the dependance of the
dispersion curves on the mobility of free charges and membrane electrical
potential. Because the $\alpha $ effect on the conductivity is much smaller
than on the dielectric constant, we are enforced to consider two different
values of the parameters: one for which one can see in detail the behavior
of dielectric constant and one for which one can see in detail the behavior
of conductivity. The results are presented in figures 2-13. The following
parameters were kept constant in our numerical analysis to the following
values (the units are in S.I): $\varepsilon _{0}=78\varepsilon _{vac}$, $%
\varepsilon _{1}=10\varepsilon _{vac}$, $\varepsilon _{2}=50\varepsilon
_{vac}$, $\sigma _{0}=0.2$, $\sigma _{1}=0$, $\sigma _{2}=0.3$, $p=0.06$, $%
\delta =1.004$.

\section{Conclusions}

We want to point out first the relation of our results with the existent
results. In the $\beta $-dispersion range of frequencies, the importance of $%
\hat{E}$ operator, in this type of analysis, was pointed out in \cite{Vr1}.
Basically, this method reduces the original equations of the electric field
to surface equations. Mathematically, this means that it transforms a
problem involving unbounded operators (Laplace operator corresponding to
certain boundary conditions) to a problem involving compact operators (the
operator $\hat{E}$). Here and in \cite{Gh2} it is given a semi-empirical
justification of the formulas (28) but, unfortunately, the formula of
polarization is not exact. Our paper states in very precise terms what
equivalation means, proves rigorously the formulas (28) and that of
polarization. Another result directly related with our approach would be 
\cite{As2}, where the method of equivalation is discussed. Note that this
result (and the method itself) depends on very particular surfaces
considered there. In the $\alpha $-dispersion range of frequencies, there
are some results for the case of spherical geometry \cite{Gh1} (and what
follows after) but, as is stated in the beginnig of this paper, the results
are valid only in a quasi-statical regime. One can see that, in the limit of
zero diffusion constants, its passing equations does not reduce to the usual
equation, but rather to two separate equations: one for conduction current
and one for displacement current. In fact, this was our reason for we have
treated very carefully the introductory sections.

In conclusion, this approach gives a unitary treatment of $\alpha $ and $%
\beta $-dispersion in the general context of arbitrary geometries. In the $%
\beta $-dispersion range of frequency, our paper corrects and puts some
existent results in rigorous settings. In the $\alpha $ range, it proposes a
completely new approach which allows us to avoid problems like that
mentioned before. Also, it proves the general formula of polarization which
includes the effects of the free charges. Our model reproduces,
qualitatively, the dielectric behavior of living cell suspensions in both $%
\alpha $ and $\beta $ frequencies ranges and is flexible enough to reproduce
any given curve which has this shape. The numerical application shows a very
strong dependency of the dielectric dispersion curves on the membrane
electrical potential. This fact anticipates the possibility of the
experimental measurement of the membrane electrical potential by a simple
measurement of the dielectric permittivity. Besides this important cell
parameter, one can measure the mobility of the free charges and, as in \cite
{PR1}, the volume concentration of the suspension. We emphasize that this
method is a non-destructive and can be a very fast one (a discussion about
this can be found in \cite{PR3}).

\newpage

{\huge Figure captions}

\bigskip

\textbf{Fig. 1} The surfaces chosen for our numerical application:

$r\left( \theta \right) =\left( 1+a\,\cos \theta \right) /\left( 1+a\right) $%
, where $a=1,2,3$.

\textbf{Fig. 2} The dependance of dielectric permittivity on membrane
potential. The numerical values are $a=1$, $D=10^{-8}$, $\gamma /D=1$
(plus), $\gamma /D=3$ (circle) and $\gamma /D=5$ (star).

\textbf{Fig. 3} The dependance of conductivity on membrane potential. The
numerical values are $a=1$, $D=10^{-8}$, $\gamma /D=1$ (plus), $\gamma /D=3$
(circle) and $\gamma /D=5$ (star).

\textbf{Fig. 4} The dependance of dielectric permittivity on membrane
potential. The numerical values are $a=1$, $D=10^{-8}$, $\gamma /D=0.1$
(plus), $\gamma /D=0.5$ (circle) and $\gamma /D=1$ (star).

\textbf{Fig. 5} The dependance of conductivity on membrane potential. The
numerical values are $a=1$, $D=10^{-8}$, $\gamma /D=0.1$ (plus), $\gamma
/D=0.5$ (circle) and $\gamma /D=1$ (star).

\textbf{Fig. 6} The dependance of dielectric permittivity on mobility. The
numerical values are $a=1$, $\gamma /D=5$, $D=10^{-7}$ (plus), $D=10^{-8}$
(circle) and $D=10^{-9}$ (star).

\textbf{Fig. 7} The dependance of conductivity on mobility. The numerical
values are $a=1$, $\gamma /D=5$, $D=10^{-7}$ (plus), $D=10^{-8}$ (circle)
and $D=10^{-9}$ (star).

\textbf{Fig. 8} The dependance of dielectric permittivity on mobility. The
numerical values are $a=1$, $\gamma /D=0.5$, $D=10^{-7}$ (plus), $D=10^{-8}$
(circle) and $D=10^{-9}$ (star).

\textbf{Fig. 9} The dependance of conductivity on mobility. The numerical
values are $a=1$, $\gamma /D=0.5$, $D=10^{-7}$ (plus), $D=10^{-8}$ (circle)
and $D=10^{-9}$ (star).

\textbf{Fig. 10} The dependance of dielectric permittivity on geometry. The
numerical values are $\gamma /D=0.5$, $D=10^{-8}$, $a=1$ (plus), $a=2$
(circle) and $a=3$ (star).

\textbf{Fig. 11} The dependance of conductivity on geometry. The numerical
values are $\gamma /D=0.5$, $D=10^{-8}$, $a=1$ (plus), $a=2$ (circle) and $%
a=3$ (star).

\textbf{Fig. 12} The dependance of dielectric permittivity on geometry. The
numerical values are $\gamma /D=5$, $D=10^{-8}$, $a=1$ (plus), $a=2$
(circle) and $a=3$ (star).

\textbf{Fig. 13} The dependance of conductivity on geometry. The numerical
values are $\gamma /D=5$, $D=10^{-8}$, $a=1$ (plus), $a=2$ (circle) and $a=3$
(star).

\FRAME{itbpF}{3.1704in}{3.1704in}{0in}{}{}{fig1.gif}{\special{language
"Scientific Word";type "GRAPHIC";display "USEDEF";valid_file "F";width
3.1704in;height 3.1704in;depth 0in;original-width 3.1254in;original-height
3.1254in;cropleft "0";croptop "1";cropright "1";cropbottom "0";filename
'Fig1.gif';file-properties "XNPEU";}}

\FRAME{itbpF}{3.1704in}{3.1704in}{0in}{}{}{fig2.gif}{\special{language
"Scientific Word";type "GRAPHIC";display "USEDEF";valid_file "F";width
3.1704in;height 3.1704in;depth 0in;original-width 3.1254in;original-height
3.1254in;cropleft "0";croptop "1";cropright "1";cropbottom "0";filename
'Fig2.gif';file-properties "XNPEU";}}

\FRAME{itbpF}{3.1704in}{3.1704in}{0in}{}{}{fig3.gif}{\special{language
"Scientific Word";type "GRAPHIC";display "USEDEF";valid_file "F";width
3.1704in;height 3.1704in;depth 0in;original-width 3.1254in;original-height
3.1254in;cropleft "0";croptop "1";cropright "1";cropbottom "0";filename
'Fig3.gif';file-properties "XNPEU";}}

\FRAME{itbpF}{3.1704in}{3.1704in}{0in}{}{}{fig4.gif}{\special{language
"Scientific Word";type "GRAPHIC";display "USEDEF";valid_file "F";width
3.1704in;height 3.1704in;depth 0in;original-width 3.1254in;original-height
3.1254in;cropleft "0";croptop "1";cropright "1";cropbottom "0";filename
'Fig4.gif';file-properties "XNPEU";}}

\FRAME{itbpF}{3.1704in}{3.1704in}{0in}{}{}{fig5.gif}{\special{language
"Scientific Word";type "GRAPHIC";display "USEDEF";valid_file "F";width
3.1704in;height 3.1704in;depth 0in;original-width 3.1254in;original-height
3.1254in;cropleft "0";croptop "1";cropright "1";cropbottom "0";filename
'Fig5.gif';file-properties "XNPEU";}}

\FRAME{itbpF}{3.1704in}{3.1704in}{0in}{}{}{fig6.gif}{\special{language
"Scientific Word";type "GRAPHIC";display "USEDEF";valid_file "F";width
3.1704in;height 3.1704in;depth 0in;original-width 3.1254in;original-height
3.1254in;cropleft "0";croptop "1";cropright "1";cropbottom "0";filename
'Fig6.gif';file-properties "XNPEU";}}

\FRAME{itbpF}{3.1704in}{3.1704in}{0in}{}{}{fig7.gif}{\special{language
"Scientific Word";type "GRAPHIC";display "USEDEF";valid_file "F";width
3.1704in;height 3.1704in;depth 0in;original-width 3.1254in;original-height
3.1254in;cropleft "0";croptop "1";cropright "1";cropbottom "0";filename
'Fig7.gif';file-properties "XNPEU";}}

\FRAME{itbpF}{3.1704in}{3.1704in}{0in}{}{}{fig8.gif}{\special{language
"Scientific Word";type "GRAPHIC";display "USEDEF";valid_file "F";width
3.1704in;height 3.1704in;depth 0in;original-width 3.1254in;original-height
3.1254in;cropleft "0";croptop "1";cropright "1";cropbottom "0";filename
'Fig8.gif';file-properties "XNPEU";}}

\FRAME{itbpF}{3.1704in}{3.1704in}{0in}{}{}{fig9.gif}{\special{language
"Scientific Word";type "GRAPHIC";display "USEDEF";valid_file "F";width
3.1704in;height 3.1704in;depth 0in;original-width 3.1254in;original-height
3.1254in;cropleft "0";croptop "1";cropright "1";cropbottom "0";filename
'Fig9.gif';file-properties "XNPEU";}}

\FRAME{itbpF}{3.1704in}{3.1704in}{0in}{}{}{fig10.gif}{\special{language
"Scientific Word";type "GRAPHIC";display "USEDEF";valid_file "F";width
3.1704in;height 3.1704in;depth 0in;original-width 3.1254in;original-height
3.1254in;cropleft "0";croptop "1";cropright "1";cropbottom "0";filename
'Fig10.gif';file-properties "XNPEU";}}

\FRAME{itbpF}{3.1704in}{3.1704in}{0in}{}{}{fig11.gif}{\special{language
"Scientific Word";type "GRAPHIC";display "USEDEF";valid_file "F";width
3.1704in;height 3.1704in;depth 0in;original-width 3.1254in;original-height
3.1254in;cropleft "0";croptop "1";cropright "1";cropbottom "0";filename
'Fig11.gif';file-properties "XNPEU";}}

\FRAME{itbpF}{3.1704in}{3.1704in}{0in}{}{}{fig12.gif}{\special{language
"Scientific Word";type "GRAPHIC";display "USEDEF";valid_file "F";width
3.1704in;height 3.1704in;depth 0in;original-width 3.1254in;original-height
3.1254in;cropleft "0";croptop "1";cropright "1";cropbottom "0";filename
'Fig12.gif';file-properties "XNPEU";}}

\FRAME{itbpF}{3.1704in}{3.1704in}{0in}{}{}{fig13.gif}{\special{language
"Scientific Word";type "GRAPHIC";display "USEDEF";valid_file "F";width
3.1704in;height 3.1704in;depth 0in;original-width 3.1254in;original-height
3.1254in;cropleft "0";croptop "1";cropright "1";cropbottom "0";filename
'Fig13.gif';file-properties "XNPEU";}}


\begin{thebibliography}{99}
\bibitem{As1}  Asami K and Yonezawa T 1996 \textit{Bioelectrochem. and
Bioenergetics} \textbf{40} 141-145

\bibitem{As2}  Asami K 1980 \textit{Japanese J. Appl. Phys} \textbf{19} 359

\bibitem{Be}  Bergman D 1978 \textit{Phys.Rep.} \textbf{43} 379

\bibitem{Bo}  Bone S and Zaba B 1992 \textit{Bioelectronics} (Chichester:
John Wiley \& Sons)

\bibitem{Gh1}  Gheorghiu E 1994\ J. \textit{Phys. A: Math. Gen.} \textbf{27}
3883

\bibitem{Gh2}  Gheorghiu E 1996 \textit{Bioelectrochem. and Bioenergetics} 
\textbf{40} 133

\bibitem{Gh3}  Gheorghiu E and Asami K 1998 \textit{Bioelectrochem. and
Bioenergetics} \textbf{45} 139

\bibitem{PR3}  Gheorghiu E, Prodan E, Mihai C and Mehedintu M 1996 \textit{%
Proceedings of IMSTEC'96} (Sydney)

\bibitem{Gi}  Gimsa J and Wachner D 1998 \textit{Biophys}. J. \textbf{75}
1107

\bibitem{Gi2}  Gimsa J, Glaser R and Fuhr G 1991 \textit{Theory and
Application of the Rotation of Biological Cells in Rotating Electric Fields
(electrorotation), In Physical Characterization of Biological Cells}
(Berlin: Verlag Gesundheit)

\bibitem{Gr}  Grosse C and Schwan H P 1992 \textit{Biophys. J}. \textbf{63}
1632

\bibitem{Pa}  Pauly H and Schwan H P 1959 \textit{Z. Naturforsh} \textbf{14b}
125

\bibitem{PR1}  Prodan E, Gheorghiu E and Vranceanu D 1996 \textit{Proc. New
Trends in Biotechnologies} (Bucharest)

\bibitem{PR2}  Prodan E 1998 \textit{J Phys A: Math Gen} \textbf{31} 4289

\bibitem{RS}  Reed M and Simon B 1978 \textit{Methods of Modern Mathematical
Physics} vol 4 (New York: Academic Press)

\bibitem{Sch}  Schwan H P 1963 \textit{Physical techniques in Biological
Research} vol 6 p 323

\bibitem{Sc}  Schanne O F, Ruiz P and Ceretti E\textit{\ 1978 Impedance
Measurements in Biological Cells} (New York, Chichester, Brisbane, Toronto:
J. Wiley\&Sons)

\bibitem{Vr1}  Vrinceanu D and Gheorghiu E 1996 \textit{Bioelectrochem. and
Bioenergetics} \textbf{40} 167
\end{thebibliography}
\end{document}